# Ferromagnetic structurally disordered ZnO implanted with Co ions

K. Potzger,[1] [(a)] Shengqiang Zhou,[1] Qingyu Xu,[1,2] A. Shalimov,[1] R. Groetzschel,[1] H. Schmidt,[1] A. Mücklich,[1] M. Helm,[1] and J. Fassbender[1]

[1]Institut für Ionenstrahlphysik und Materialforschung, Forschungszentrum Dresden-Rossendorf e.V., Bautzner Landstraße 128, 01328 Dresden, Germany
[2]Department of Physics, Southeastern University, Nanjing 211189, China

We present superparamagnetic clusters of structurally highly disordered Co–Zn–O created by high fluence Co ion implantation into ZnO (0001) single crystals at low temperatures. This secondary phase cannot be detected by common x-ray diffraction but is observed by high-resolution transmission electron microscopy. In contrast to many other secondary phases in a ZnO matrix, it induces low-field anomalous Hall effect and is thus a candidate for magnetoelectronics applications. ©2008 *American Institute of Physics*

## Contents

- BODY OF ARTICLE
- REFERENCES
- FIGURES
- FOOTNOTES

Recently, much effort has been undertaken for the creation of diluted magnetic semiconductors (DMSs) based on transition metal (TM) doped ZnO. Theoretically, both antiferromagnetic superexchange as well as ferromagnetic *d-d* double exchange are present in those materials.[1] Both become stronger with increasing TM to Zn ratio. It was recently proven that $Zn_{1-x}Co_xO$ with a perfect crystalline lattice is basically paramagnetic with evidence for antiferromagnetic coupling.[2] Concordantly, we suggest the enhancement of the Co to Zn ratio along with the creation of crystalline disorder and nanostructures for ferromagnetism in Co implanted ZnO. Bond angle disorder can, for example, weaken antiferromagnetic superexchange.[3] Experimentally, ferromagnetic hysteresis loops have been observed in several disordered/defective TM oxide nanostructures.[4,5,6] It was shown that CoO nanoparticles become ferromagnetic after annealing in nitrogen.[4] Dutta *et al.*[4] considered uncompensated spins at the surface to be only a negligible contribution to the room-temperature ferromagnetism of the greenish-turquoise CoO nanoparticles. The existence of ferromagnetic Co–Zn–O crystalline/amorphous core-shell nanostructures has been shown in Ref. 5. A detailed investigation on the ferromagnetic properties of disordered NiO was given by Yi *et al.*[6] Supposing that disordered Co–Zn–O also exhibits ferromagnetic properties, we created it by Co ion implantation into hydrothermal ZnO (0001) single crystals from Crystec. We used an implantation energy of 80 keV and an angle of 7° in order to avoid channeling effects. The target holder was cooled by liquid nitrogen during implantation. We have chosen two Co fluences, i.e., $8\times10^{16}$ and $16\times10^{16}$ $Co^+/cm^2$ [the corresponding samples are further called ZnO(8) and ZnO(16)]. Such large fluences are expected to heavily disorder the crystal structure.[7] Nevertheless, after implantation the crystals appear greenish and transparent.

The formation of a nanoscale secondary phase from this large Co concentration can lead to superparamagnetism. Exactly such behavior was observed by means of superconducting quantum interference device (SQUID) (QD MPMS XL) magnetometry. Figure [1](a) shows zero-field-cooled/field-cooled (ZFC/FC) magnetization versus temperature curves obtained at a magnetic field of 100 Oe applied parallel to the sample surface. As expected, the thermomagnetic irreversibility temperature increases with increasing fluence. Ferromagnetic hysteresis loops with a saturation magnetization of $0.05\mu_B$/Co ion for ZnO(8) (not shown) and $0.35\mu_B$/Co ion for ZnO(16) [Fig. [1](b)] were observed. For the calculation of the magnetization, we took into account the resputtered Co ions with a percentage of 19% and 43% for ZnO(8) and ZnO(16), respectively. Those values were calculated using TRIDYN (Ref. [8]) (see also below). The experimental data on magnetization (Fig. [1]) are modelled using a Preisach approach (red solid curves) described by Song et al.[9] The temperature dependence of the parameters $p$ describing the magnetic properties of the (nano)phase is usually expressed by the critical temperature $T_C$ and the critical exponent $\Gamma$: $p=p_0(1-T/T_C)^{\Gamma}$, with $p$ substituted by the mean magnetic moment $\mu$ of an individual cluster, the mean coercivity $H_c$, or the dispersions $\sigma_i$ and $\sigma_c$ of the interparticle interaction and coercive fields. All parameters $p_0$ as well as $T_C$ increase with increasing fluence: $\mu_0$ ($10.000\mu_B \rightarrow 150.000\mu_B$), $H_{c0}$ (100 Oe $\rightarrow$ 200 Oe), $\sigma_{i0}$ (150 Oe $\rightarrow$ 200 Oe), $\sigma_{c0}$ (250 Oe $\rightarrow$ 400 Oe), and $T_C$ (250 K $\rightarrow$ 600 K). Due to coalescence, the total number of clusters $N$ with respect to a 1 $cm^2$ area implanted decreases from $52\times10^{10}$ to $25\times10^{10}$. Although the cluster moments found by the Preisach model are somewhat idealized, an assumption of a Co spin moment of $3\mu_B$ for both samples roughly suggests a particle diameter increase by a factor of 2.5. The presence of large ferromagnetic domains for ZnO(16) is also evident from atomic/magnetic force microscopy (AFM/MFM). The micrographs [Figs. [1](c)[1](d)] were recorded with a Veeco/DI Multimode. Magnetic contrast shows an array of domains with sizes of 0.5,…,1 $\mu$m, which partially follow topography. It was recorded at room temperature in remanent state. Note the peculiarity that the ZnO surface contains self-arranged hillocks created due to ion impact [Fig. [1](c)]. Their origin is not the ion beam erosion but rather corresponds to the decomposition of ZnO, as observed during vacuum annealing of ZnO substrates.[10]

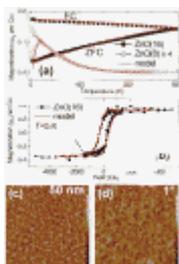
[Figure 1.](#)

Although superparamagnetic clusters are present, $\theta$-$2\theta$ scans using a Siemens D5000 x-ray diffractometer in Bragg–Brentano as well as $2\theta$ scans in glancing incidence geometry revealed no crystalline phases other than wurtzite ZnO to be present in ZnO(16) (not shown). Metallic crystalline Co nanoparticles would, on the other hand, be easily detectable[11] at a high amount of $2\times10^{16}$ $Co^0/cm^2$, i.e., 2 $\mu g/cm^2$ of metallic Co with a magnetization of $1.6\mu_B$/atom (again the resputtered Co ions are taken into account). Thus, the superparamagnetic nanoclusters are either wurtzite with similar lattice parameters as ZnO or structurally highly disordered. The latter can be expected from the large fluence implanted. Structural disorder due to ion implantation in ZnO was described in detail by Kucheyev *et al.*[7] At Au fluences as large as $3\times10^{16}$ cm$^{-2}$, the ZnO crystals are heavily disordered, but no amorphous phase was detected. Concerning the energy used in Ref. 7 and the different masses of Au and Co, the maximum total displacements per atom are half as for our ZnO(16). For a structure and composition analysis of both samples, we performed Rutherford backscattering/channeling (RBS/C) experiments using a primary energy of 1.7 MeV. Transmission electron microscopy (TEM) using a FEI Titan machine equipped with an energy dispersive x-ray spectrometer (EDXS) was applied only to ZnO(16) (Fig. 2). The main observations using these methods are as follows.

• First, heavy disorder of the crystalline matrix is reflected by a broad bulk damage peak extending from 50 to 150 nm below the surface [Fig. 2(a)]. The bulk channeling reflects the maximum disorder in the crystals and is similar (saturated) for both fluences. As in Ref. 7, a damage peak close to the surface can be detected (dotted arrow). As observed by TEM, the implanted film close to the surface [Fig. 2(b)] is a mixture of a crystalline dominated area and an amorphous dominated area. The crystalline dominated area is basically wurtzite but structurally disordered due to dislocations and precipitations (see below). The amorphous phase can be observed either by focusing on a very thin part of the specimen [Fig. 2(b)] or as rings in the fast Fourier transform (FFT) [Fig. 2(c)]. The rings appear under appropriate defocusing of the electron beam. They are not present for crystals with a long range order, such as single crystalline Si. Note that the amorphous dominated area contains low crystalline cores of only ~4 nm size, which is nearly not dispersed. They basically exhibit a similar crystalline orientation like the substrate, and they are surrounded by interconnected amorphous shells [Fig. 2(b)].

• Second, the maximum Co to Zn relation for ZnO(16) indicated by the full arrow in Fig. 2(a) and derived from grazing incidence RBS (not shown) amounts to 0.33 (10). The maximum ratio between Co and Zn derived by the TRIDYN (Ref. 8) code amounts to 0.37 for ZnO(8) and 0.62 for ZnO(16). This value is significantly larger than the one found by RBS. This implicates diffusion of Co toward the ZnO bulk. The latter was confirmed by using EDXS [inset in Fig. 2(b)]. The thickness of the Co$^+$ implanted layer approximated by the full width at half maximum amounts to 75 nm as compared to 45 nm as derived by TRIDYN.

• Third, stoichiometric disorder is indicated. EDXS point measurements at distances of 25 and 50 nm from the surface (not shown) reveal two kinds of O:(Zn+Co) ratios. The first equals unity proofing Zn and Co to be mainly in the 2+ state. The second—which occurs rather sporadically—exceeds unity significantly. Correspondingly, nanoscale regions with crystalline superstructure [FFT in Fig. 2(d)] sporadically occur. Their net-plane distances of 0.290 and 0.242 nm can be assigned to the (220) and (311) Miller indices of Zn(or Co)Co$_2$O$_4$. They nearly correspond to those of the bulk materials.[12] We found fixed orientation relationships of [111] ∥ ZnO[001] and (110) ∥ ZnO(110) as well as [111] ∥ ZnO[001] and (112) ∥ ZnO(110). Streaklike shapes of the spots indicate a distribution of the lattice parameters.

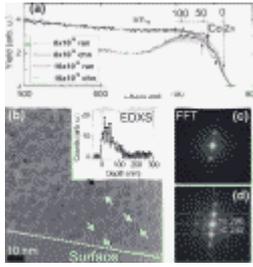
Figure 2.

The surface layer of the otherwise insulating ZnO single crystals becomes *n*-type conducting after implantation. In our case, ZnO(16) exhibits anomalous Hall effect (AHE), as shown in Fig. 3. The saturation field amounts to ~6 kOe for 5 K. It persists up to room temperature (not shown) and implicates charge carrier spin polarization. The magnetoresistance (MR) effect is negative, amounting to −0.15% at 10 kOe and 5 K (left inset in Fig. 3). The room-temperature resistivity of ZnO(16) amounts to $2.4\times10^{-3}$ $\Omega$ cm if a film thickness of 75 nm is assumed. For a metallic Co film, it is much lower, i.e., $6\times10^{-6}$ $\Omega$ cm.[13] The latter rules out the presence of a pure metallic film. The temperature dependence of the sheet resistance $R(s)$ (Fig. 3, right inset) hints to different transport mechanisms inside the material with mostly semiconducting behavior, indicating a multiphase system. Especially, at around 80 K there is a pronounced change in the slope. The room-temperature charge carrier concentration of ZnO(16) is $1.3\times10^{21}$ cm$^{-3}$, assuming a thickness of 75 nm.

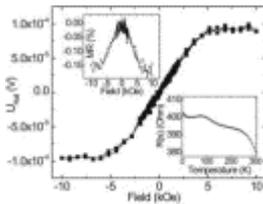
Figure 3.

In conclusion, high fluence Co$^+$ ion implantation into ZnO single crystals leads to the formation of superparamagnetic clusters consisting of Zn, Co, and O. We relate them to nanosized regions consisting of low crystalline cores of 4 nm diameter and interconnected amorphous shells created by the ion beam impact. This secondary phase occurs simultaneously with areas of wurtzite as well as spinel structures. Crystalline $Zn_{1-x}Co_xO$ and $Zn(or Co)Co_2O_4$ cannot account for the pronounced ferromagnetic signal since the former is paramagnetic and the latter occurs only sporadically. Thus, relating the structural to the magnetic properties, a likely mechanism is ferromagnetic 3*d* exchange fostered by the crystalline disorder. Suppression of antiferromagnetic coupling, nanoscale dimensions of the clusters, or stoichiometric disorder might be relevant to the magnetic ordering. In the future it has to be investigated whether fully amorphous regions or such with residual crystalline relations are responsible for the signal. As an outlook, following Ref. 6, a continuous fully amorphous phase cannot account for ferromagnetism in NiO. Moreover, for disordered ferromagnetic phases in other DMSs,[14] also an increase in magnetization with the formation of crystalline/amorphous nanostructures has been found. The implanted sheet shows pronounced AHE, which is saturated at low field, giving hope for applicability in spintronics.

A.S. wants to thank the Deutsche Forschungsgemeinschaft (DFG) (Project No. PO1275/2-1, "SEMAN" and S.Z., Q.X., and H.S. would like to thank the Bundesministerium für Bildung und Forschung (Grant No. FKZ03N8708) for funding.

## REFERENCES

Citation links [e.g., Phys. Rev. D 40, 2172 (1989)] go to online journal abstracts. Other links (see Reference Information) are available with your current login. Navigation of links may be more efficient using a second browser window.